\documentstyle[preprint,aps,epsfig]{revtex}
\tightenlines
\begin{document}
\draft

\title{Virtual photon fragmentation functions}
\author{Jianwei Qiu and Xiaofei Zhang}
\address{Department of Physics and Astronomy,
         Iowa State University \\
         Ames, Iowa 50011, USA}

\date{February 9, 2001}
\maketitle
\begin{abstract}
We introduce operator definitions for virtual photon fragmentation
functions, which are needed for reliable calculations of Drell-Yan
transverse momentum ($Q_T$) distributions when $Q_T$ is much larger
than the invariant mass $Q$.  We derive the evolution equations for
these fragmentation functions.  We calculate the leading order
evolution kernels for partons to fragment into a unpolarized as well
as a polarized virtual photon.  We find that fragmentation functions
to a longitudinally polarized virtual photon are most important at
small $z$, and the fragmentation functions to a transversely polarized 
virtual photon dominate the large $z$ region.  We discuss the
implications of this finding to the J/$\psi$ mesons' polarization at
large transverse momentum.  
\end{abstract}
\vspace{0.2in}


\section{Introduction}

Gluon distribution plays a central role in calculating many important
signatures at hadron colliders because of the dominance of gluon
initiated subprocesses.  A precise knowledge of the gluon distribution
is absolutely vital for reliable predictions for signal as well as
background cross sections \cite{CTEQ-G}.  A great effort has been
devoted to find good physical observables for extracting information
on gluon distribution \cite{G-pdf,QCD-rpt}.  

For many years, prompt photon production has been thought as a clean
signal for information 
on gluon distribution because its cross section at lowest order is
dominated by the ``Compton'' subprocess: $g+q\rightarrow \gamma+q$,
and this dominance is preserved at higher orders
\cite{Owens-RMP,BOO-ph,BQ-Photon,Aurenche}. However, the theoretical
and experimental complications have limited our ability to extract
clean information on gluon distribution from the direct photon data
\cite{CTEQ5}.  At collider energies, prompt photons are observed and
their cross sections are measured only if the photons are relatively
isolated in phase space.  Isolation is required to reduce various
hadronic backgrounds. But at the same time, the cross section is no
longer totally inclusive and theoretical predictions become sensitive
to the isolation parameters \cite{BQ-Photon,BGQ-iso}.  In addition, 
phenomenological fragmentation functions are needed for
including photons emerged from the long-distance fragmentation of
quarks and gluons that are themselves produced at short-distance
\cite{Owens-RMP,Aurenche}.  Our knowledge on phenomenological
fragmentation functions and the theoretical uncertainties associated
with the isolation prevent fully quantitative determinations of gluon
distribution from the collider data on isolated photons.  
Although data at fixed target energies provide good
information on gluon distribution at large $x$ \cite{MRST}, the
controversy about how much $k_T$-smearing is required to understand 
the data introduces the significant uncertainties to the gluon
distributions \cite{CTEQ-kt,Aurenche2}.  
As a result, all direct photon data were excluded from recent CTEQ
global analyses of parton distributions \cite{CTEQ5}.    

Recently, Berger, Gordon, and Klasen (BGK) show that Drell-Yan
transverse momentum ($Q_T$) distributions in hadronic 
collisions are dominated by partonic subprocesses initiated by
incident gluons if $Q_T > Q/2$, where $Q$ is the invariant mass of the
produced lepton pairs \cite{BGK-DY}.  BGK argue that
Drell-Yan $Q_T$ distribution is an advantageous
source of constraints on the gluon distribution, free from the
experimental and theoretical complications of photon isolation that
beset studies of prompt photon production.

Other than the difference between a virtual and a real photon,
Drell-Yan and prompt photon production share the same partonic  
subprocesses.  The virtual photon produced in Drell-Yan process
subsequently decays into a pair of leptons.  Since we are mainly
interested in the 
cross section at high $Q_T$ and low $Q$, we ignore the $Z$ channel
contributions in this paper.  If we integrate over angular dependence
of the lepton pairs, the Drell-Yan massive lepton-pair production
between hadron $A$ and $B$ can be expressed in terms of an inclusive
production of a virtual photon \cite{BGK-DY} 
\begin{equation}
\frac{d\sigma_{AB\rightarrow \ell^+\ell^-(Q)}}{dQ^2\,dQ_T^2\,dy}
= \left(\frac{\alpha_{em}}{3\pi Q^2}\right)
  \frac{d\sigma_{AB\rightarrow \gamma^*(Q)}}{dQ_T^2\,dy}\, .
\label{DY-Vph}
\end{equation}
Because of the advantage of measuring the leptons, Drell-Yan
massive lepton-pair production as well as the inclusive virtual photon
production defined in Eq.~(\ref{DY-Vph}) are entirely inclusive. 
The usual factorization theorems in Quantum Chromodynamics (QCD) 
should apply \cite{CSS-fac,Bodwin},  
\begin{equation}
\frac{d\sigma_{AB\rightarrow \gamma^*(Q)}}{dQ_T^2\,dy}
=\sum_{a,b}\int dx_1 \phi_{a/A}(x_1,\mu) 
           \int dx_2 \phi_{b/B}(x_2,\mu)\,
 \frac{d\hat{\sigma}_{ab\rightarrow \gamma^*(Q)}}{dQ_T^2\,dy}
 (x_1,x_2,Q,Q_T,y;\mu)
\label{Vph-fac}
\end{equation}
where $\sum_{a,b}$ run over all parton flavors, the $\phi_{a/A}$
and $\phi_{b/B}$ are parton distributions, and $\mu$ represents both
renormalization and factorization scale.  In Eq.~(\ref{Vph-fac}),  
$d\hat{\sigma}_{ab\rightarrow \gamma^*(Q)}/dQ_T^2 dy$ are
short-distance partonic hard parts and perturbatively calculable to
all orders in $\alpha_s(\mu)$.  Similar to prompt photon
production, the lowest order ``Compton'' subprocess to a virtual
photon: $g+q\rightarrow \gamma^*+q$ dominates the $Q_T$ distributions
at large $Q_T$ as long as the collision energy is high enough to
overcome the phase space penalty caused by the virtual photon mass.  
Therefore, Drell-Yan $Q_T$ distribution at large $Q_T$ is an
advantageous source of information on the gluon distribution
\cite{BGK-DY}. 

When $Q_T\gg Q$, the $Q_T$ distributions calculated order-by-order in
$\alpha_s$ in the conventional fixed-order QCD perturbation theory
receive a large logarithm, $\ln(Q_T^2/Q^2)$, at every power of
$\alpha_s$ beyond the leading order.  Therefore, at sufficiently large
$Q_T$ and $\sqrt{S}$, the convergence of the conventional perturbative
expansion in powers of $\alpha_s$ is impaired, and the logarithms must
be resummed. 

In order to resum the large logarithm, we introduce a concept of
virtual photon fragmentation functions 
$D_{f\rightarrow \gamma^*}(z,\mu_F;Q)$ for 
a parton of flavor $f$ to fragment into a virtual photon of invariant 
mass $Q$.  Normally, a virtual particle state is not physical, and
therefore, a fragmentation function to such a state may be gauge
dependent and ill-defined.  However, if such a virtual state
immediately decays into a {\it completely} measured physical state, we  
believe that a fragmentation function to such a virtual state is
effectively physical.  The fragmentation function is experimentally
measurable if the decay to the physical state is calculable. 

Unlike the real photon fragmentation functions \cite{Owens-RMP}, the
virtual photon fragmentation functions are fully perturbative if $Q\gg
\Lambda_{\rm QCD}$.  In terms of the virtual photon fragmentation
functions, the conventional perturbative expansion for  
$d\hat{\sigma}_{ab\rightarrow \gamma^*(Q)}/dQ_T^2\,dy$ in
Eq.~(\ref{Vph-fac}) can be {\it reorganized} according to a new 
factorization formula such that the large logarithms are resummed to
all orders in $\alpha_s$.  The detailed derivation of the new
factorization formula for the virtual photon production at $Q_T\ge Q$
and the discussions on the all order resummations will be published
elsewhere.  In this paper, we concentrate on the process
independent physics associated with the virtual photon fragmentation
functions. 

In the next section, we derive the cut-vertex and corresponding
operator definitions for virtual photon fragmentation functions.  
From the definitions, we derive the evolution equations (or
renormalization group equations) for these fragmentation functions. We
calculate the leading order evolution kernels for the evolution
equations.  By solving the evolution equations with the calculated
evolution kernels, we derive the virtual photon fragmentation
functions.  We show that the virtual photon fragmentation functions
are in principle perturbatively calculable to all orders in
$\alpha_s$ if $Q\gg \Lambda_{\rm QCD}$. 

Because of our ability of measuring the leptons, we can probe the
polarization of the virtual photon in Drell-Yan massive lepton-pair
production.  Therefore, it is also physically meaningful to define
fragmentation functions to a virtual photon with a specific
polarization.  In Sec.~\ref{sec3}, we derive the evolution equations
and the leading order evolution kernels to polarized virtual
photon fragmentation functions.  

In Sec.~\ref{sec4}, we present our numerical results for
virtual photon fragmentation functions.  By showing the fragmentation
functions at different scales, we demonstrate the evolution properties
of the fragmentation functions for partons to a unpolarized as well as
a polarized virtual photon.  We find that fragmentation
functions to a longitudinally polarized virtual photon are most
important at small $z$, and the fragmentation functions to a
transversely polarized virtual photon dominate the large $z$ region.
When $Q_T$ is large while $\sqrt{S}$ is
fixed, fragmentation functions at large $z$ are more relevant for
calculating the cross sections.  Therefore, we conclude that the
virtual photons produced in a unpolarized Drell-Yan massive
lepton-pair production are more likely to be transversely polarized at
high $Q_T$.  

Recent data on J/$\psi$ polarization measured by CDF collaboration at
Fermilab Tevatron seem to be inconsistent with the predictions from
various models of J/$\psi$ production \cite{CDF-jpsi}.  
The virtual photon production (extracted from Drell-Yan massive
lepton-pair production) at large $Q_T$ and small $Q^2$ has a lot in
common with the J/$\psi$ production at high $Q_T$.  They both have two
large physical scales: $Q_T$ and $Q^2$, which is equal to $M_{{\rm
J/}\psi}^2$ in the case of J/$\psi$ production; and $Q_T^2$ is much
larger than $Q^2$.  If the collision energy $\sqrt{S}$ is large enough
and the logarithm $\ln(Q_T^2/Q^2)$ is so important that the resummed
fragmentation contributions dominate the production cross sections,
the virtual photon and J/$\psi$ production will share the {\it same}
partonic subprocesses.  Only difference between the virtual photon and
J/$\psi$ production at high $Q_T$ is the fragmentation functions.  
The virtual photon fragmentation functions are completely
perturbative, while the parton to J/$\psi$ fragmentation functions
involve final-state nonperturbative interactions.  Understanding the
difference in such final-state interactions is very important for
reliable predictions of J/$\psi$ production.  We propose to
measure the virtual photon polarization in Drell-Yan massive
lepton-pair production at large $Q_T$ and low $Q^2$.  Because the
virtual photon polarizations in Drell-Yan massive lepton-pair
production are completely calculable and independent of the
final-state nonperturbative effect, the measurements of 
the virtual photon polarizations at high $Q_T$ provide not only a
good test of QCD perturbation theory, but also a reference process to
test the models of J/$\psi$ formation.  

\section{Unpolarized virtual photon fragmentation functions}
\label{sec2}

Like other fragmentation functions \cite{CS-pdf}, a virtual photon
fragmentation function, $D_{f\rightarrow \gamma^*}(z,\mu_F;Q)$ is
defined as a probability density to find a virtual photon of invariant
mass $Q$ and momentum fraction $z$ from a parent parton of flavor $f$.
In this section, we derive the operator definitions for virtual photon
fragmentation functions, and corresponding evolution equations.  We
calculate the leading order evolution kernels for the evolution
equations and derive the virtual photon fragmentation functions by
solving the evolution equations numerically.

In order to simply our derivations, we choose a frame in which the
virtual photon is moving very fast along the $z$-axis, 
\begin{equation}
Q^{\mu}=(Q^+,Q^-,0_T),  \quad \mbox{and} \quad
Q^-=\frac{Q^2}{2Q^+}
\label{Q-vec} \\
\end{equation}
with $Q^+\gg Q^-$. We also introduce two useful vectors
\begin{equation}
\bar{n}^{\mu}=(1,0,0_T) \quad 
\mbox{and} \quad
n^{\mu}=(0,1,0_T)
\label{nbar-n}
\end{equation}
with $\bar{n}^2=n^2=0$ and $\bar{n}^\mu n_\mu=1$.  For any four-vector
$p$, we have $p^\mu \bar{n}_\mu = p^-$ and $p^\mu n_\mu = p^+$.

\begin{figure}
\begin{center}
\epsfig{figure=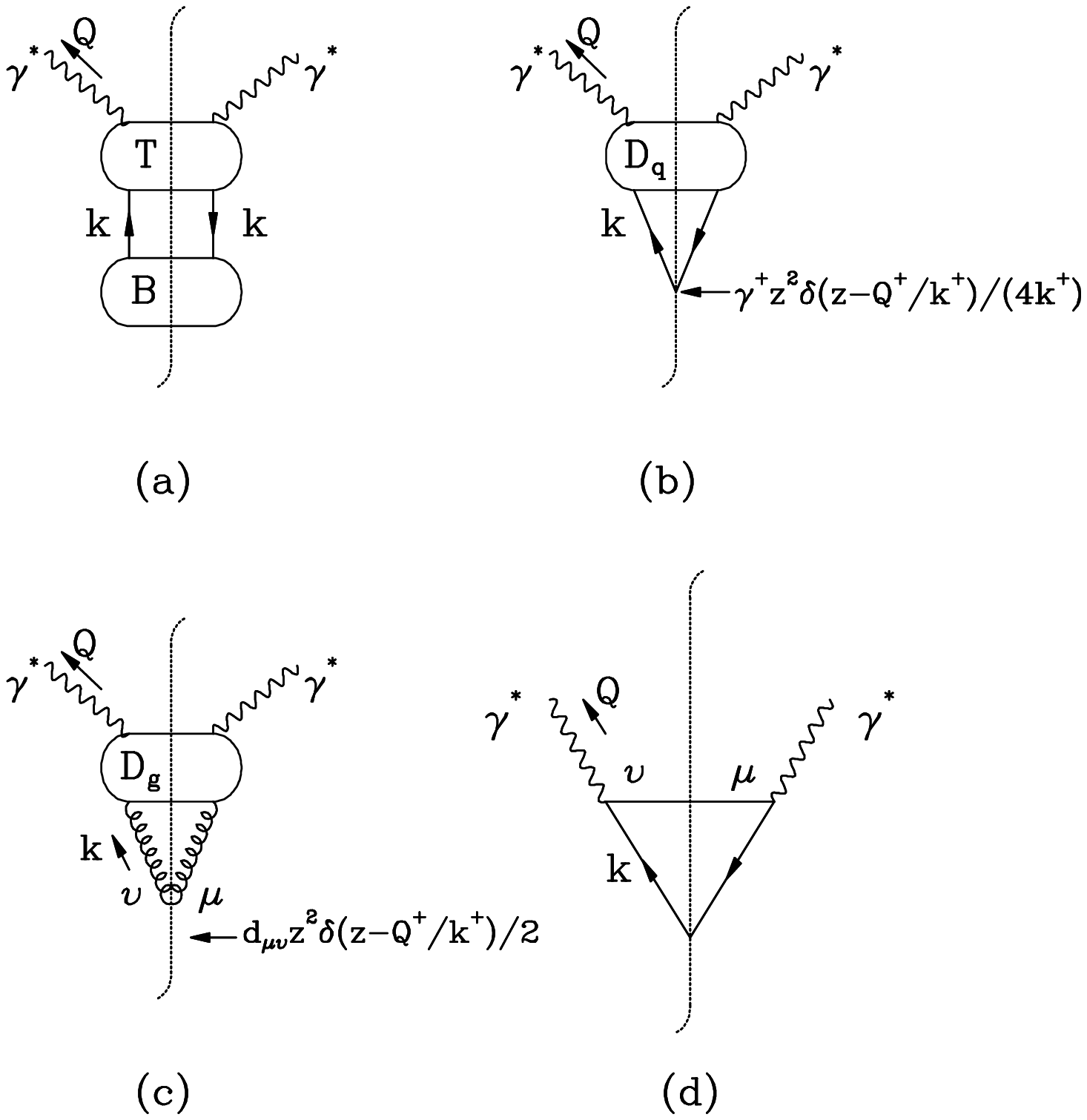,width=5.0in}
\end{center}
\caption{(a) A generic diagram for a scattering process in which a
quark of momentum $k$ fragments into a virtual photon of invariant
mass $Q$; (b) the cut-vertex diagram for quark-to-virtual-photon
fragmentation function; (c) the cut-vertex diagram for
gluon-to-virtual-photon fragmentation function; (d) the lowest order
cut-vertex diagram for a quark to a virtual photon.}
\label{fig1}
\end{figure}

Consider a generic quark to virtual photon fragmentation process, as
shown in Fig.~\ref{fig1}(a).  The top part, labeled by $T$,
corresponds to the fragmentation from a quark of momentum $k$ to a
virtual photon of invariant mass $Q$; and the bottom part, labeled by
$B$, represents a short-distance hard collision at a scale $\mu \gg
Q$.  By carrying out collinear expansion of the quark momentum $k$
in the $B$ at $k^\mu = (Q^+/z)\bar{n}^\mu$ and separation of the trace
between the top and the bottom \cite{Qiu-T4}, we factorize the generic
fragmentation process as follows,
\begin{eqnarray}
&&\int \frac{d^4k}{(2\pi)^4}\, {\rm Tr}\left[
 B_q(\mu,k)\, T_{q\rightarrow\gamma^*}(k,Q)\right]
\nonumber \\
&\ & {\hskip 0.5in} \approx
\int \frac{dz}{z^2}\,
{\rm Tr}\left[B_q(\mu,k=\frac{Q^+}{z})
   \left\{\gamma^-\left(\frac{Q^+}{z}\right)\right\} \right] 
\nonumber \\
&\ & {\hskip 0.8in} \times
\int \frac{d^4k}{(2\pi)^4}\, {\rm Tr}\left[
\left\{\frac{\gamma^+}{4k^+}\,z^2\, \delta(z-\frac{Q^+}{k^+})\right\} 
              T_{q\rightarrow\gamma^*}(k,Q)\right] 
\nonumber \\
&\ & {\hskip 0.5in} \equiv
\int \frac{dz}{z^2}\, 
H_q(\mu,k=\frac{Q^+}{z})\,
D_{q\rightarrow\gamma^*}(z,\mu_F;Q)\, .
\label{q-fac}
\end{eqnarray}
where $H_q(\mu,k=Q^+/z)\equiv
{\rm Tr}[B_q(\mu,k=Q^+/z)\{\gamma^-(Q^+/z)\}]$ represents the leading
power short-distance production of a quark of momentum
$k^\mu=(Q^+/z)\bar{n}^\mu$ and  
\begin{equation}
D_{q\rightarrow\gamma^*}(z,\mu_F;Q)
\equiv \int \frac{d^4k}{(2\pi)^4}\, {\rm Tr}\left[
\left\{\frac{\gamma^+}{4k^+}\,z^2\, \delta(z-\frac{Q^+}{k^+})\right\} 
              T_{q\rightarrow\gamma^*}(k,Q)\right] 
\label{Dq-CutV}
\end{equation}
is the quark-to-virtual-photon fragmentation function in terms of its
cut-vertex definition \cite{AHM-CutV}. As shown in Fig.~\ref{fig1}(b),  
the $\{\frac{\gamma^+}{4k^+}\,z^2\, \delta(z-\frac{Q^+}{k^+})\}$ in
Eq.~(\ref{Dq-CutV}) is the corresponding cut-vertex.  The cut-vertex
for an antiquark-to-virtual-photon fragmentation function is the same
as that for a quark-to-virtual-photon fragmentation function. 

Similarly, by considering a generic gluon to virtual photon
fragmentation process, we derive the cut-vertex definition for the
gluon-to-virtual-photon fragmentation function with the cut-vertex
$\{\frac{1}{2}\, d^{\mu\nu}\,z^2\, \delta(z-\frac{Q^+}{k^+})\}$, as
shown in Fig.~\ref{fig1}(c).  The tensor $d^{\mu\nu}$ is defined as 
\begin{equation}
d^{\mu\nu} = 
-g^{\mu\nu} + \bar{n}^\mu\, n^\nu + n^\mu\, \bar{n}^\nu \, .
\label{dmn}
\end{equation}

From the cut-vertex definitions in Figs.~\ref{fig1}(b) and
\ref{fig1}(c), we derive corresponding operator definitions for
virtual photon fragmentation functions as follows.  By representing
the diagram in Fig.~\ref{fig1}(b) in terms of quark fields, we have
\begin{eqnarray}
D_{q\rightarrow \gamma^*}(z,\mu_F;Q) &=&
\int \frac{d^4k}{(2\pi)^4}\left[ z^2\ \delta(z-\frac{Q^+}{k^+})\,
\frac{1}{4k^+}\right]
(2\pi)^4\delta^4(k-Q-\sum_X k_X) 
\prod_X \frac{d^3k_X}{(2\pi)^3 2E_X}
\nonumber\\
&\ & {\hskip 0.2in} \times 
\frac{1}{N}\sum_{i=1}^{N}
{\rm Tr}\left[\gamma^+ 
\langle 0|\psi_{q_i}(0)|\gamma^*(Q)X\rangle
\langle X\gamma^*(Q)|\bar{\psi}_{q_i}(0)|0\rangle \right]
\label{Dq-OP} \\
&=&
\frac{z}{4} \int \frac{dy^-}{2\pi}\,
{\rm e}^{-i(Q^+/z)y^-}\, 
\frac{1}{N}\sum_{i=1}^{N}
{\rm Tr}\left[\gamma^+ 
\langle 0|\psi_{q_i}(0)|\gamma^*(Q)\rangle
\langle \gamma^*(Q)|\bar{\psi}_{q_i}(y^-)|0\rangle \right]\, .
\nonumber
\end{eqnarray}
where $(1/N)\sum_{i=1}^{N}$ with $N=3$ indicates the average over the 
quark color.  Similarly, we derive the operator definition for the
gluon-to-virtual-photon fragmentation function,
\begin{eqnarray}
D_{g\rightarrow \gamma^*}(z,\mu_F;Q) &=&
\frac{z^2}{2Q^+} \int \frac{dy^-}{2\pi}\,
\mbox{e}^{-i(q^+/z)y^-}\, (-g_{\mu\nu})
\nonumber \\
&\times & 
\frac{1}{N^2-1}\sum_{a=1}^{N^2-1}
\langle 0|F^{+\mu}_a(0)|\gamma^*(Q)\rangle
\langle \gamma^*(Q)|F^{+\nu}_a(y^-)|0\rangle\, .
\label{Dg-OP}
\end{eqnarray}
Both operator definitions in Eqs.~(\ref{Dq-OP}) and (\ref{Dg-OP}) 
are in the light-cone gauge.  Proper insertion of a line integral 
of the color and electromagnetic potential is needed to make them both
color and electromagnetic gauge invariant \cite{CS-pdf,Braaten}.  

From either the cut-vertex definitions or the operator definitions,
we derive the evolution equations (or renormalization group equations)
for virtual photon fragmentation functions \cite{CQ-evo}, 
\begin{eqnarray}
\mu_F^2 \frac{d}{\mu_F^2} D_{c\rightarrow\gamma^*}(z,\mu_F;Q)
&=& 
\left(\frac{\alpha_{em}}{2\pi}\right)
 \gamma_{c\rightarrow\gamma^*}(z,\mu_F,\alpha_s;Q)
\nonumber \\
&+&
\left(\frac{\alpha_{s}}{2\pi}\right)
\sum_d \int_z^1 \frac{dz'}{z'}
P_{c\rightarrow d}(\frac{z}{z'},\alpha_s)\, 
 D_{d\rightarrow\gamma^*}(z',\mu_F;Q)\, ,
\label{RG-unpol}
\end{eqnarray}
where $c,d=q,\bar{q},g$.  The fragmentation scale $\mu_F$ is 
defined to be the scale where the operators of the fragmentation
functions are renormalized.  We choose the $\mu_F^2$ to be the
invariant mass of the fragmenting parton at the renormalization
point.  

The evolution kernels $\gamma_{c\rightarrow\gamma^*}$ and 
$P_{c\rightarrow d}$ in Eq.~(\ref{RG-unpol}) have the following
perturbative expansions,
\begin{eqnarray}
\gamma_{c\rightarrow\gamma^*}(z,\mu_F,\alpha_s;Q)
&=& \sum_{n=0} \gamma_{c\rightarrow\gamma^*}^{(n)}(z,\mu_F;Q)
               \left(\frac{\alpha_s}{2\pi}\right)^n\, ,
\label{Evo-ph}\\
P_{c\rightarrow d}(z,\alpha_s)
&=& \sum_{n=0} P_{c\rightarrow d}^{(n)}(z)
              \left(\frac{\alpha_s}{2\pi}\right)^n\, ,
\label{Evo-ap}
\end{eqnarray}
where the renormalization scale dependence is suppressed. 
In our derivation of Eq.~(\ref{RG-unpol}), we assume a strong ordering
in partons' invariant mass in the fragmentation sequence, except the
last step (quark-to-virtual photon).  As a result, 
the evolution kernels $P_{c\rightarrow d}$ for the homogeneous terms
in Eq.~(\ref{RG-unpol}) are the same as the kernels for the DGLAP
equation \cite{CTEQ-book}.  Although the evolution equations in
Eq.~(\ref{RG-unpol}) have the same functional form as that for real
photon fragmentation functions \cite{Owens-RMP}, the evolution kernels
$\gamma_{c\rightarrow\gamma^*}$ are different due to the nonvanish
mass of the virtual photon.

In order to calculate the evolution kernels
$\gamma_{c\rightarrow\gamma^*}$, we need to specify the polarization
vector $\epsilon^\mu_\lambda(Q)$ for the virtual photon state
$|\gamma^*(Q)\rangle$.  For a unpolarized virtual photon, we need only
the following polarization tensor,
\begin{equation}
P^{\mu\nu}(Q) \equiv \sum_{\lambda=T,L} 
\epsilon^{*\mu}_{\lambda}(Q)\,
\epsilon^{\nu}_{\lambda}(Q)\, ,
\label{Pol-unpol}
\end{equation}
where $T$ and $L$ represent the virtual photon's transverse and
longitudinal polarization, respectively.  Although the evolution
kernels and the fragmentation functions defined in
Eqs.~(\ref{Dq-OP}) and (\ref{Dg-OP}) are gauge invariant,  
the functional form of the polarization tensor $P^{\mu\nu}(Q)$ as well
as the number of Feynman diagrams contributing to the evolution
kernels are gauge dependent.   

In the light-cone gauge, we have the polarization tensor
\begin{equation}
P^{\mu\nu}(Q) 
=-g^{\mu\nu} + \frac{Q^\mu n^\nu + n^\mu Q^\nu}{Q\cdot n}\, ,
\label{Pol-unpol-lcg}
\end{equation}
and have only one Feynman diagram, as shown in Fig.~\ref{fig1}(d),
which contributes to the lowest order quark-to-virtual-photon
fragmentation function.  We obtain
\begin{equation}
D_{q\rightarrow \gamma^*}^{(0)}(z,\mu_F;Q) 
= e_q^2 \left(\frac{\alpha_{em}}{2\pi}\right)
      \Bigg[ \left(\frac{1+(1-z)^2}{z}\right)
             \ln\left(\frac{z\mu_F^2}{Q^2}\right)
            -z\left(1-\frac{Q^2}{z\mu_F^2}\right) \Bigg] ,
\label{Dq0-unpol}
\end{equation}
where $e_q$ is the fractional charge for the quark of flavor $q$, and
the fragmentation scale $\mu_F^2\ge Q^2/z$ due to the kinematics.  

\begin{figure}
\begin{center}
\epsfig{figure=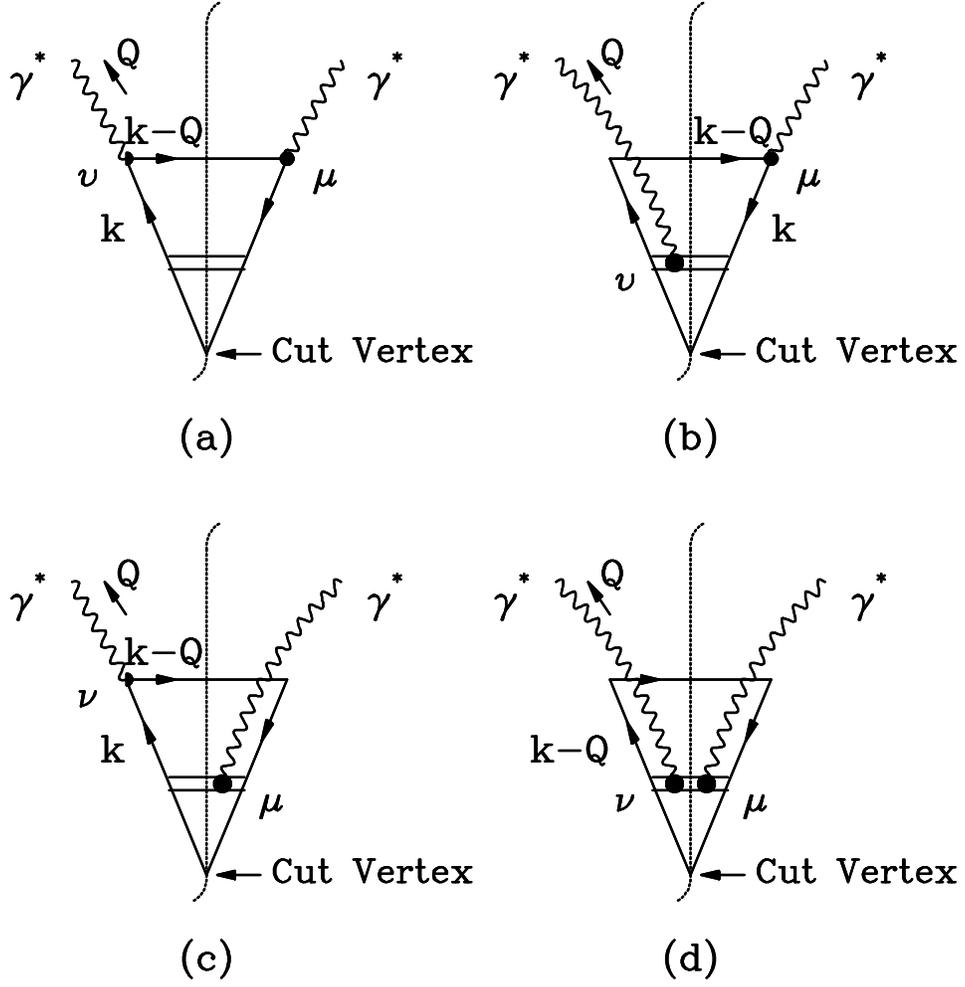,width=5.0in}
\end{center}
\caption{Lowest order gauge invariant set of Feynman diagrams for
quark-to-virtual-photon fragmentation functions
\protect\cite{BQ-spin}.  } 
\label{fig2}
\end{figure}

For an arbitrary choice of gauge, we need a total of four Feynman
diagrams for the lowest order quark-to-virtual-photon fragmentation,
as shown in Fig.~\ref{fig2} \cite{BQ-spin}.  The 
diagrams in Fig.~\ref{fig2} contain double ``eikonal'' lines, and the
Feynman rules for the double ``eikonal'' lines can be found in
Refs.~\cite{CS-pdf,BQ-spin}.  These four diagrams in Fig.~\ref{fig2}
form a gauge invariant set at this order \cite{BQ-spin,BL-DY}.
Contributions of the three diagrams from 
Figs.~\ref{fig2}(b) to \ref{fig2}(d) vanish when they are contracted
by the light-cone gauge polarization tensor in
Eq.~(\ref{Pol-unpol-lcg}).  With the covariant polarization tensor
\begin{equation}
P^{\mu\nu}(Q) 
=-g^{\mu\nu} + \frac{Q^\mu Q^\nu}{Q^2}\, ,
\label{Pol-unpol-cg}
\end{equation}
or simply $P^{\mu\nu}(Q) = -g^{\mu\nu}$, we obtain the same lowest
order quark-to-virtual-photon fragmentation function in
Eq.~(\ref{Dq0-unpol}).  Since gluon does not directly couple to a
photon, we have  
\begin{equation}
D_{g\rightarrow \gamma^*}^{(0)}(z,\mu_F;Q) = 0\, .
\label{Dg0-unpol} 
\end{equation}

By applying $\mu_F^2 d/d\mu_F^2$ to the lowest order virtual photon
fragmentation functions, we obtain the lowest order evolution kernels
\begin{eqnarray}
\gamma_{q\rightarrow \gamma^*}^{(0)}(z,\mu_F;Q) 
&=&
e_q^2 \Bigg[\frac{1+(1-z)^2}{z}
           -z\left(\frac{Q^2}{z\mu_F^2}\right) \Bigg]
      \theta(\mu_F^2-\frac{Q^2}{z})\, ,
\nonumber \\
\gamma_{g\rightarrow \gamma^*}^{(0)}(z,\mu_F;Q) 
&=& 0\, .
\label{G0-unpol} 
\end{eqnarray}
With the lowest order evolution kernels $\gamma_{q\rightarrow
\gamma^*}^{(0)}$ in Eq.~(\ref{G0-unpol}) and 
$P_{c\rightarrow d}^{(0)}$ from Ref.~\cite{CTEQ-book}, 
we solve the evolution equations in Eq.~(\ref{RG-unpol}) 
for the unpolarized virtual photon fragmentation functions, and 
present the numerical results in Sec.~\ref{sec4}.
 
\section{Polarized virtual photon fragmentation functions}
\label{sec3}

By measuring the momenta of both leptons in Drell-Yan massive
lepton-pair production, we can determine both invariant mass and
polarization state of the virtual photon, which decays into the
lepton-pair.  Therefore, it is meaningful to define the fragmentation
functions to a polarized virtual photon.  In this section, we
calculate the evolution kernels for the evolution equations of
polarized virtual photon fragmentation functions.

Since we did not use any constraints due to the virtual photon's
polarization when we derived the cut-vertex and operator definitions
for the virtual photon fragmentation functions in last section, the
same definitions should be valid for fragmentation functions to a
polarized virtual photon.  Therefore, with a proper choice of
polarization vectors $\epsilon^\mu_\lambda(Q)$ for a virtual photon of
polarization $\lambda$, we can calculate the fragmentation functions
to a virtual photon of a specific polarization state.

In order to specify a polarized virtual photon state, we need to
define the photon's polarization vector $\epsilon^\mu_\lambda(Q)$.
If the photon is real ($Q^2=0$), it has
only {\it two} transverse polarization states and its longitudinal
polarization can be gauged away by an extra gauge transformation.
When the photon is virtual and its quantum numbers are completely
fixed by the physical observables, the extra gauge degree of freedom
used to remove the longitudinal polarization at $Q^2 = 0$ is lost, 
and therefore, the virtual photon has three polarization states. 

With our choice of the frame in which the virtual photon is moving
very fast along the $z$-axis, we define the
photon's polarization vectors as the following,
\begin{eqnarray}
\epsilon^{\mu}_{T=1}(Q)&=&(0,1,0,0) \, ,
\nonumber \\
\epsilon^{\mu}_{T=2}(Q)&=&(0,0,1,0) \, ,
\nonumber \\
\epsilon^{\mu}_{L}(Q)&=&\frac{1}{\sqrt{2}}
         ([Q^+-Q^-],0,0,[Q^++Q^-]) \, ,
\label{pol-vec}
\end{eqnarray}
which are effectively the same as those in the ``S-helicity'' frame
defined in Ref.~\cite{LT-DY}.  We obtain corresponding
polarization tensors,  
\begin{equation}
P^{\mu\nu}_T(Q) \equiv \frac{1}{2}\ \sum_{T=1,2} 
\epsilon^{*\mu}_{T}(Q)\,
\epsilon^{\nu}_{T}(Q)
=\frac{1}{2}\, d^{\mu\nu} 
\label{Pol-T} 
\end{equation}
with $d^{\mu\nu}$ defined in Eq.~(\ref{dmn}) and 
\begin{equation}
P_{L}^{\mu\nu}(Q) \equiv 
\epsilon^{*\mu}_{L}(Q)\,
\epsilon^{\nu}_{L}(Q)
\label{Pol-L}
\end{equation}
for transversely and longitudinally polarized virtual photons,
respectively.  By summing over all polarization states, we should
recover the polarization tensor for a unpolarized virtual photon,
\begin{equation}
P^{\mu\nu}(Q) = 2\  P^{\mu\nu}_T(Q) +  P^{\mu\nu}_L(Q) \, ,
\label{Pol-sum} 
\end{equation}
where the factor of two represents the virtual photon's two transverse
polarization states.

Applying the transverse polarization tensor $P_T^{\mu\nu}$ to 
the lowest order Feynman diagrams in
Fig.~\ref{fig2}, we obtain the lowest order fragmentation function
for a quark to a transversely polarized virtual photon as
\begin{equation}
D_{q\rightarrow \gamma_T^*}^{(0)}(z,\mu_F;Q) 
= e_q^2 \left(\frac{\alpha_{em}}{2\pi}\right)\, 
\frac{1}{2}\, \left(\frac{1+(1-z)^2}{z}\right)
\left[\ln\left(\frac{z\mu_F^2}{Q^2}\right)
       -\left(1-\frac{Q^2}{z\mu_F^2}\right) \right]
\label{Dq0-T}
\end{equation}
with $\mu_F^2\ge Q^2/z$.
Again, the lowest order fragmentation function for a gluon to a
photon vanishes, $D_{g\rightarrow \gamma_T^*}^{(0)}(z,\mu_F;Q)=0$. 
Corresponding evolution kernels are given by
\begin{eqnarray}
\gamma_{q\rightarrow \gamma_T^*}^{(0)}(z,\mu_F;Q) 
&=& e_q^2\, \frac{1}{2}\, 
\left(\frac{1+(1-z)^2}{z}\right)
 \left[1-\frac{Q^2}{z\mu_F^2}\right]
  \theta(\mu_F^2-\frac{Q^2}{z}) \, ,
\nonumber \\
\gamma_{g\rightarrow \gamma_T^*}^{(0)}(z,\mu_F;Q) 
&=& 0\, .
\label{G0-T} 
\end{eqnarray}
As expected, when $Q\rightarrow 0$,
$2\,\gamma_{q\rightarrow\gamma_T^*}^{(0)}$ in  
Eq.~(\ref{G0-T}) reduces to the lowest order evolution kernels for the
real photon fragmentation functions \cite{Owens-RMP}.  

Similarly, by applying the longitudinal polarization tensor
$P_L^{\mu\nu}$ in Eq.~(\ref{Pol-L}) to the Feynman diagrams in
Fig.~\ref{fig2}, we derive  
\begin{equation}
D_{q\rightarrow \gamma_L^*}^{(0)}(z,\mu_F;Q) 
= e_q^2 \left(\frac{\alpha_{em}}{2\pi}\right)
 \left[2\left(\frac{1-z}{z}\right)\right]
      \left(1-\frac{Q^2}{z\mu_F^2}\right)\, ,
\label{Dq0-L}
\end{equation}
with $\mu_F^2\ge Q^2/z$, 
and $D_{g\rightarrow \gamma_L^*}^{(0)}(z,\mu_F;Q)=0$.  
Our lowest order fragmentation function to a longitudinally polarized
virtual photon in Eq.~(\ref{Dq0-L}) is consistent with the result
derived by Braaten and Lee \cite{BL-DY}.  
The $(1-z)$ factor in Eq.~(\ref{Dq0-L}) is a consequence of the vector
interaction between the quark and photon.  As a consistency check, our
lowest order polarized virtual photon fragmentation functions in 
Eqs.~(\ref{Dq0-T}) and (\ref{Dq0-L}) satisfy
\begin{equation}
2\ D_{q\rightarrow \gamma_T^*}^{(0)}(z,\mu_F;Q) 
+ D_{q\rightarrow \gamma_L^*}^{(0)}(z,\mu_F;Q) 
= D_{q\rightarrow \gamma^*}^{(0)}(z,\mu_F;Q) \, ,
\label{Dq0-sum}
\end{equation}
where $D_{q\rightarrow \gamma^*}^{(0)}$ is given in
Eq.~(\ref{Dq0-unpol}). 
From Eq.~(\ref{Dq0-L}), we derive the evolution kernels for
longitudinally polarized virtual photon fragmentation functions
\begin{eqnarray}
\gamma_{q\rightarrow \gamma_L^*}^{(0)}(z,\mu_F;Q) 
&=& e_q^2 
\left[2\left(\frac{1-z}{z}\right)\right]
       \left(\frac{Q^2}{z\mu_F^2}\right)
  \theta(\mu_F^2-\frac{Q^2}{z})\, ,
\nonumber \\
\gamma_{g\rightarrow \gamma_L^*}^{(0)}(z,\mu_F;Q) 
&=& 0\, .
\label{G0-L} 
\end{eqnarray}
Again, we have $2\, \gamma_{q\rightarrow \gamma_T^*}^{(0)}
+ \gamma_{q\rightarrow \gamma_L^*}^{(0)} = 
\gamma_{q\rightarrow \gamma^*}^{(0)}$.  

Since the polarized evolution kernels and the polarized virtual photon
fragmentation functions are gauge invariant, we can also derive them
from the single diagram in Fig.~\ref{fig1}(d) in the light-cone gauge. 
Substituting $Q^\mu=Q^+\bar{n}^\mu + Q^- n^\mu$ into the polarization
tensor in the light-cone gauge in Eq.~(\ref{Pol-unpol-lcg}), we obtain
\begin{equation}
P^{\mu\nu}(Q) = d^{\mu\nu} + \frac{Q^2}{(Q^+)^2}\, n^\mu\, n^\nu\, .
\label{Pol-sum-lcg}
\end{equation}
Since the vector $n^\mu$ used to fix the light-cone gauge does not
have transverse component, the $d^{\mu\nu}$ in Eq.~(\ref{Pol-sum-lcg})
should still be identified as $2\, P_T^{\mu\nu}(Q)$. 
From Eq.~(\ref{Pol-sum}), we obtain the polarization tensor for a
longitudinally polarized virtual photon 
\begin{equation}
P^{\mu\nu}_L(Q) = \frac{Q^2}{(Q^+)^2}\, n^\mu\, n^\nu
\label{Pol-L-lcg}
\end{equation}
in the light-cone gauge.  Applying this $P^{\mu\nu}_L(Q)$ to the
Feynman diagram in Fig.~\ref{fig1}(d), we can derive the lowest order 
fragmentation functions to a longitudinally polarized virtual photon.
As expected, the derived fragmentation functions are the same as those
given in Eq.~(\ref{Dq0-L}). 

Since the fragmentation functions to a polarized and a unpolarized
virtual photon share the same form of the operator definitions,  
the evolution equations for polarized virtual photon fragmentation
functions have the same functional form as that in
Eq.~(\ref{RG-unpol}), except the evolution kernels
$\gamma_{c\rightarrow\gamma^*}$ are replaced by those in
Eq.~(\ref{G0-T}) and Eq.~(\ref{G0-L}) for transversely polarized and
longitudinally polarized virtual photon, respectively.  
Since the evolution kernels $P_{c\rightarrow d}$ are independent of
the polarization of the produced virtual photon, they should remain
the same.

\section{Numerical Results and Conclusions}
\label{sec4}

In last two sections, we derived the analytical expressions for the
lowest order parton-to-virtual-photon fragmentation functions with the
virtual photon unpolarized as well as polarized.  
The lowest order gluon-to-virtual-photon fragmentation
functions vanish while the lowest order quark-to-virtual-photon
fragmentation functions are given in Eqs.~(\ref{Dq0-unpol}),
(\ref{Dq0-T}), and (\ref{Dq0-L}) for unpolarized, transversely
polarized, and longitudinally polarized, respectively.  
Strong interactions in the top blob of the cut-vertex diagrams in
Figs.~\ref{fig1}(b) and \ref{fig1}(c) can change 
the $z$ as well as the $\mu_F$ dependence of the lowest order virtual
photon fragmentation functions.  Solving the evolution equations 
in Eq.~(\ref{RG-unpol}) is effectively to resum all leading
logarithmic contributions from the strong interactions.  The power
corrections from the strong interactions O($\Lambda_{\rm
QCD}^2/\mu_F^2$) are not included in the evolution equations.

In order to solve the evolution equations in Eq.~(\ref{RG-unpol}), we
need to specify a boundary condition.  Unlike the real photon
fragmentation functions, we do not need any nonperturbative input
distributions if the invariant mass $Q\gg\Lambda_{\rm QCD}$.  From the
pure kinematics, we have the following boundary condition for solving
the evolution equations in Eq.~(\ref{RG-unpol}),
\begin{equation}
D_{f\rightarrow \gamma^*}(z,\mu_F^2\leq Q^2/z;Q)=0 
\label{B-cond}
\end{equation}
for all parton flavor $f=q,\bar{q},g$ and any polarization of the
virtual photon. 

Since the boundary conditions given in Eq.~(\ref{B-cond}) are the same
for all flavors of massless partons, and the evolution kernels
$\gamma_{c\rightarrow\gamma^*}$ in Eq.~(\ref{RG-unpol}) are the same
for a quark $q$ and corresponding antiquark $\bar{q}$, we
have
\begin{equation}
D_{q\rightarrow \gamma^*}(z,\mu_F;Q)
= D_{\bar{q}\rightarrow \gamma^*}(z,\mu_F;Q)
\label{q-qbar}
\end{equation}
for all quark flavor $q$.  By neglecting the quark mass difference,
the only flavor dependence of the evolution kernels
$\gamma_{c\rightarrow\gamma^*}$ in Eq.~(\ref{RG-unpol}) is from
quark's fractional charge $e_q$.  Therefore, quark-to-virtual-photon
fragmentation functions are the same for all quark flavors with the
same fractional charge.

With the boundary condition given in Eq.~(\ref{B-cond}), we can solve
the evolution equations in Eq.~(\ref{RG-unpol}) in the moment-space
analytically, and then, perform the Mellin transformation from the
moment-space back to the $z$-space \cite{Owens-RMP}.  However, 
we find that it is easier to solve the evolution equations numerically
in the $z$-space directly.

\begin{figure}
\begin{minipage}[t]{3in}
\epsfig{figure=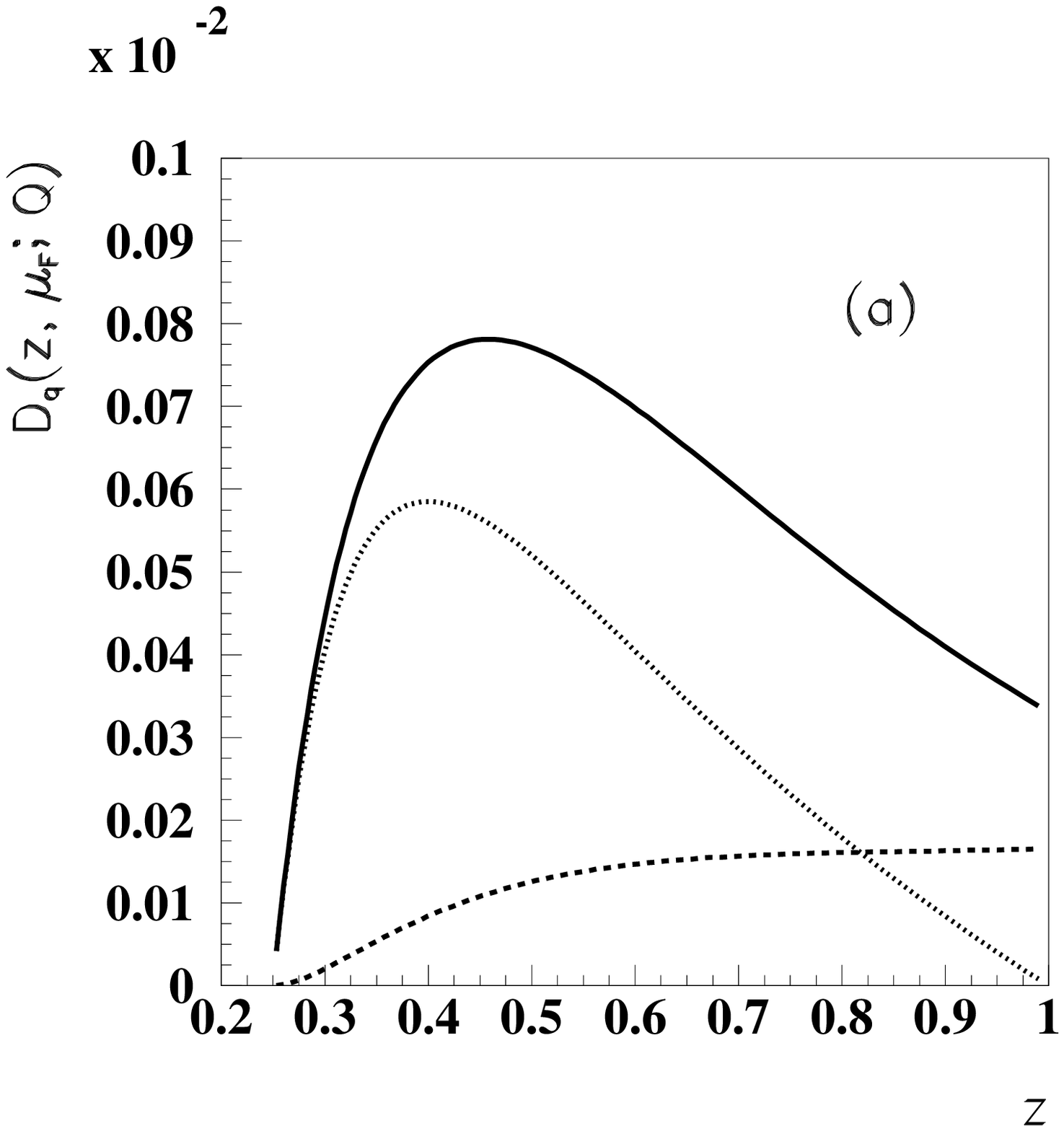,width=3.0in}
\end{minipage}
\hfil
\begin{minipage}[t]{3in}
\epsfig{figure=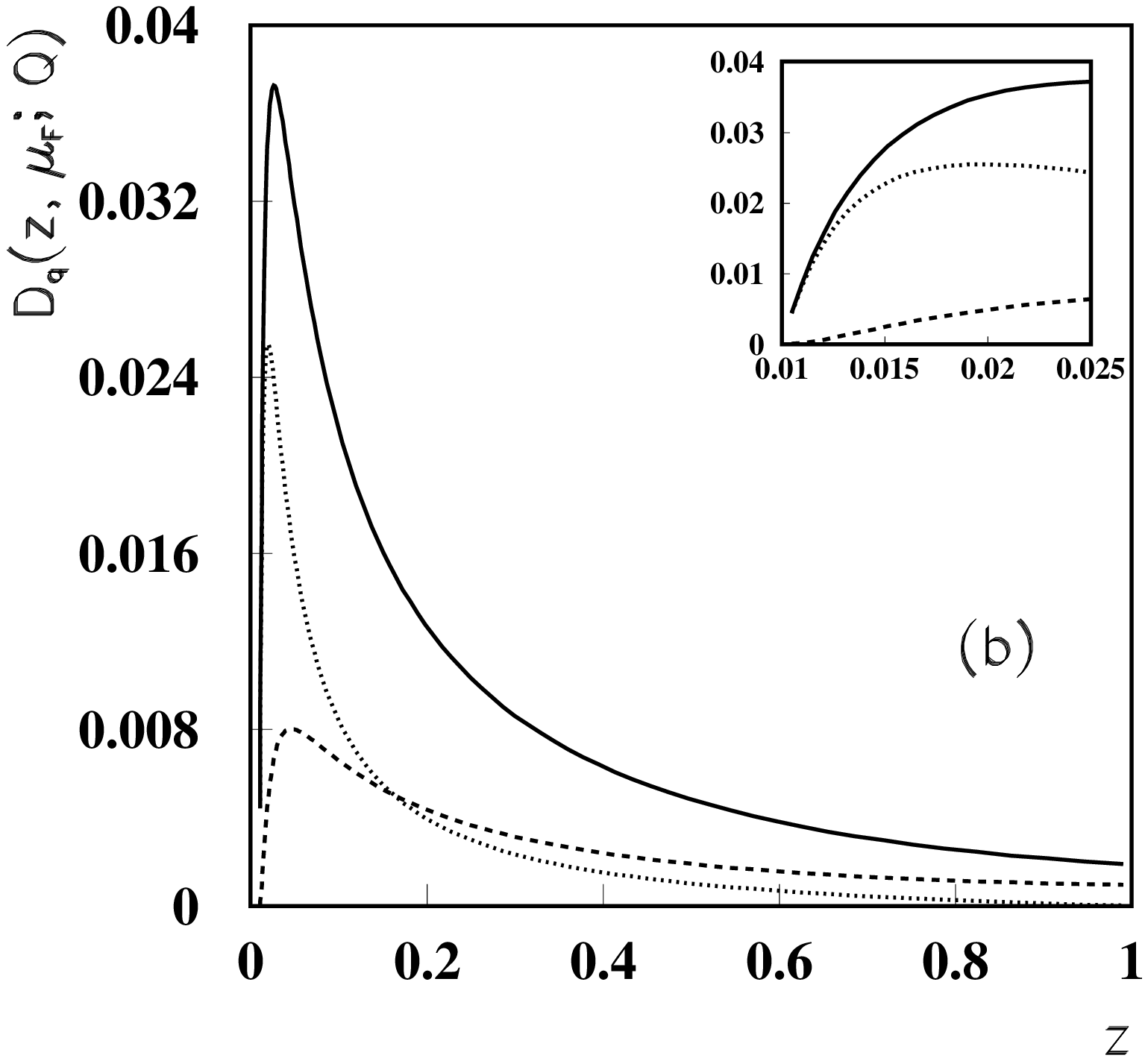,width=3.0in}
\end{minipage}
\caption{The lowest order virtual photon fragmentation functions as a
function of $z$ at $Q=5$~GeV and $\mu_F=10$~GeV (a), and
$\mu_F=50$~GeV (b).  The solid, dashed, and dotted lines are for
unpolarized, transversely polarized, and longitudinally polarized
virtual photons, respectively. The inset in Fig.~\protect\ref{fig2}(b)
shows the $z\le 0.025$ region. }  
\label{fig3}
\end{figure}

In Figs.~\ref{fig3}(a) and \ref{fig3}(b), we plot the derived lowest
order quark-to-virtual-photon fragmentation functions as a function of
the momentum fraction $z$ at a fragmentation scale $\mu_F=10$~GeV and  
$\mu_F=50$~GeV, respectively.  We choose $e_q=2/3$ for the quark's
fractional charge, and the virtual photon's invariant mass to be
$Q=5$~GeV.  The unpolarized quark-to-virtual-photon 
fragmentation functions given in Eq.~(\ref{Dq0-unpol}) are
represented by the solid lines.  The transversely and longitudinally
polarized virtual photon fragmentation functions given in
Eqs.~(\ref{Dq0-T}) and (\ref{Dq0-L}) are represented by the dashed
and dotted lines, respectively.  The solid lines are equal to twice of
the dashed lines plus the dotted lines. which is a consequence of
Eq.~(\ref{Dq0-sum}).  

From Fig.~\ref{fig3}, we find that longitudinally polarized virtual
photon fragmentation functions are much larger than the transversely
polarized virtual photon fragmentation functions when $z$ is small.
The small $z$ region corresponds to the region where $\mu_F^2$ is close
to the threshold $Q^2/z$.  Near the threshold, the evolution kernels
for the transversely polarized fragmentation functions 
$\gamma_{q\rightarrow\gamma_T^*}{(0)}$ in Eq.~(\ref{G0-T}) 
vanish while the kernels for the longitudinally polarized
fragmentation functions $\gamma_{q\rightarrow\gamma_L^*}^{(0)}$ in
Eq.~(\ref{G0-L}) are finite and large.  The 
$\gamma_{q\rightarrow\gamma_L^*}^{(0)}$ are actually proportional to
$1/z$ when $\mu_F^2\rightarrow Q^2/z$.  Therefore, the longitudinally
polarized virtual photon fragmentation functions dominate the small
$z$ or the threshold region. 

On the other hand, the fragmentation
functions for a transversely polarized virtual photon evolve much
faster than the longitudinally polarized fragmentation functions in
the large $z$ region, as shown in Fig.~\ref{fig3}.  This is because
the evolution kernels $\gamma_{q\rightarrow\gamma_L^*}^{(0)}$ 
for the longitudinally polarized virtual photon fragmentation
functions are power suppressed (proportional to $1/\mu_F^2$) and also
vanish as $z\rightarrow 1$.  

\begin{figure}
\begin{center}
\begin{minipage}[t]{3in}
\epsfig{figure=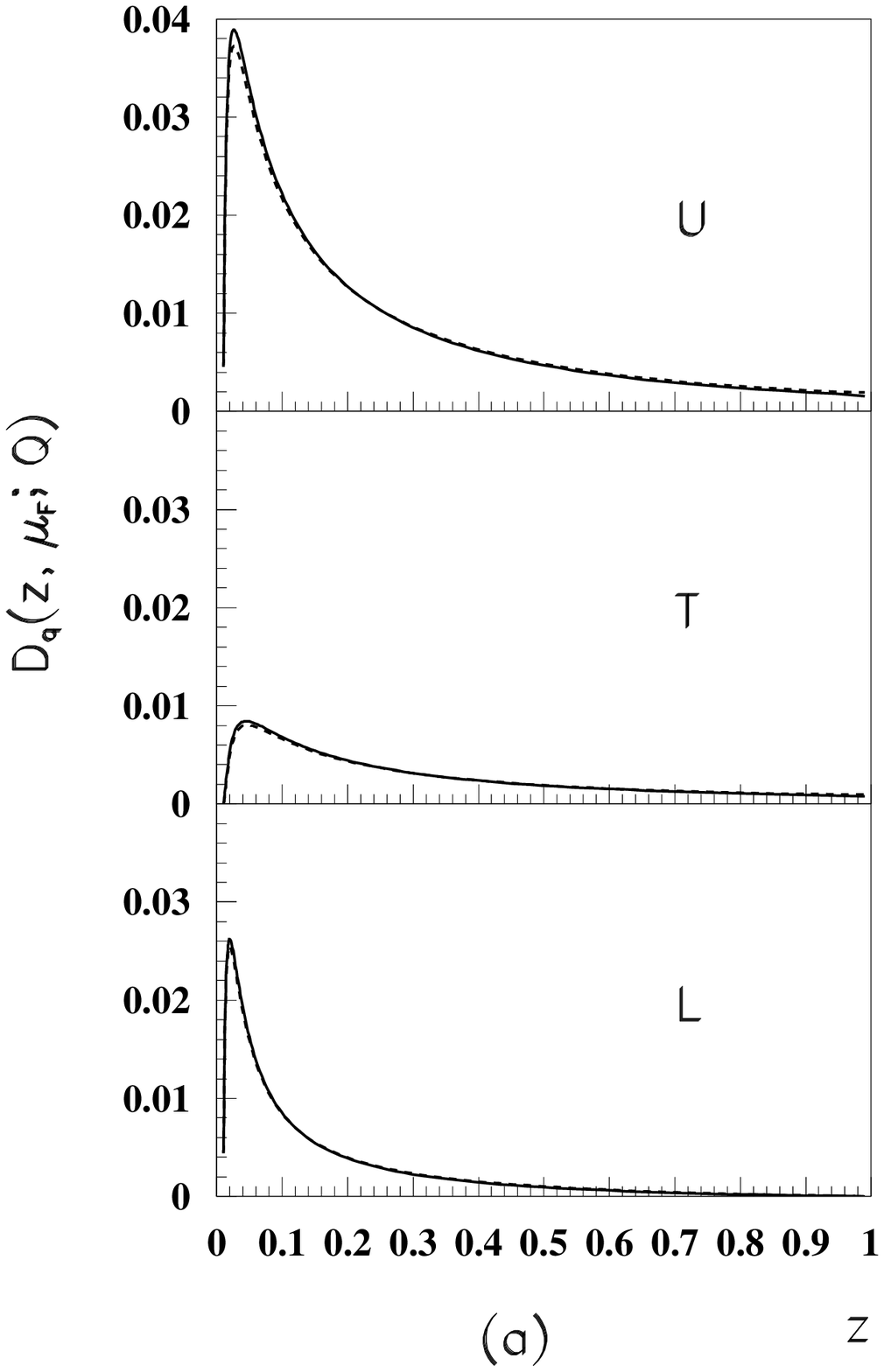,width=3.0in}
\end{minipage}
\hfil
\begin{minipage}[t]{3in}
\epsfig{figure=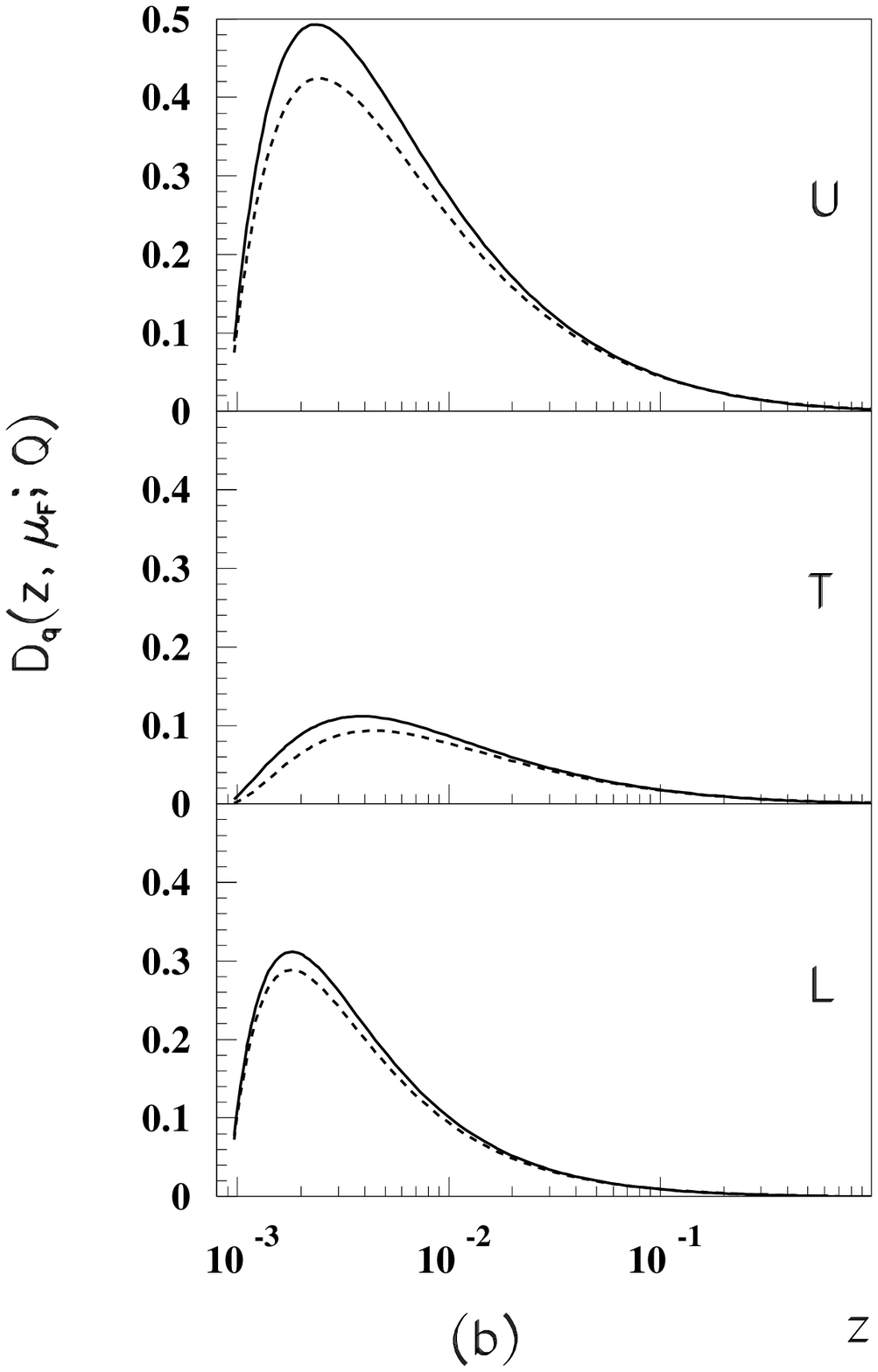,width=3.0in}
\end{minipage}
\end{center}
\caption{Comparison between the lowest order (dashed) and the QCD
evolved (solid) virtual photon fragmentation functions as a function
of $z$ at $\mu_F=50$~GeV and $Q=5$~GeV (a) and $Q=1.5$~GeV (b).}
\label{fig4}
\end{figure}

In order to see the effect of resumming the large logarithms from
quark and gluon radiations, we numerically solve the evolution
equations in Eq.~(\ref{RG-unpol}) with the lowest order evolution
kernels $\gamma_{q\rightarrow\gamma^*}^{(0)}$ and 
$P_{c\rightarrow d}^{(0)}$.  We plot the comparison between the
lowest order (dashed) and the QCD evolved (solid)
quark-to-virtual-photon fragmentation functions at $Q=5$~GeV and
$\mu_F=50$~GeV in Fig.~\ref{fig4}(a).  The symbols: U, T, and L
represent the fragmentation functions to a unpolarized, transversely
polarized, and longitudinally polarized virtual photon, respectively.  
Like the evolution of parton 
distributions, QCD evolution tries to enlarge the fragmentation
functions in the small $z$ region, and to suppress the fragmentation
functions in the large $z$ region.  The difference between the
QCD evolved and lowest order fragmentation functions at $Q=5$~GeV is
small because of the boundary conditions in Eq.~(\ref{B-cond}).
However, when $Q$ is smaller or $\mu_F$ is larger, we expect QCD
evolution to be much more important because 
of a larger logarithm $\ln(\mu_F^2/Q^2)$.  For example, cross
sections for virtual photon production ware measured by UA1
Collaboration at CERN for the virtual photon mass $Q\in [2m_\mu,
2.5]$~GeV \cite{UA1-Vph}.  Instead of $Q=5$~GeV in
Fig.~\ref{fig4}(a), we plot the same quark-to-virtual-photon
fragmentation functions at $Q=1.5$~GeV in Fig.~\ref{fig4}(b).
Clearly, QCD evolved virtual photon fragmentation functions in
Fig.~\ref{fig4}(b) are enhanced in comparison with those in 
Fig.~\ref{fig4}(a), particularly, in the small $z$ region.  

\begin{figure}
\begin{minipage}[t]{3in}
\epsfig{figure=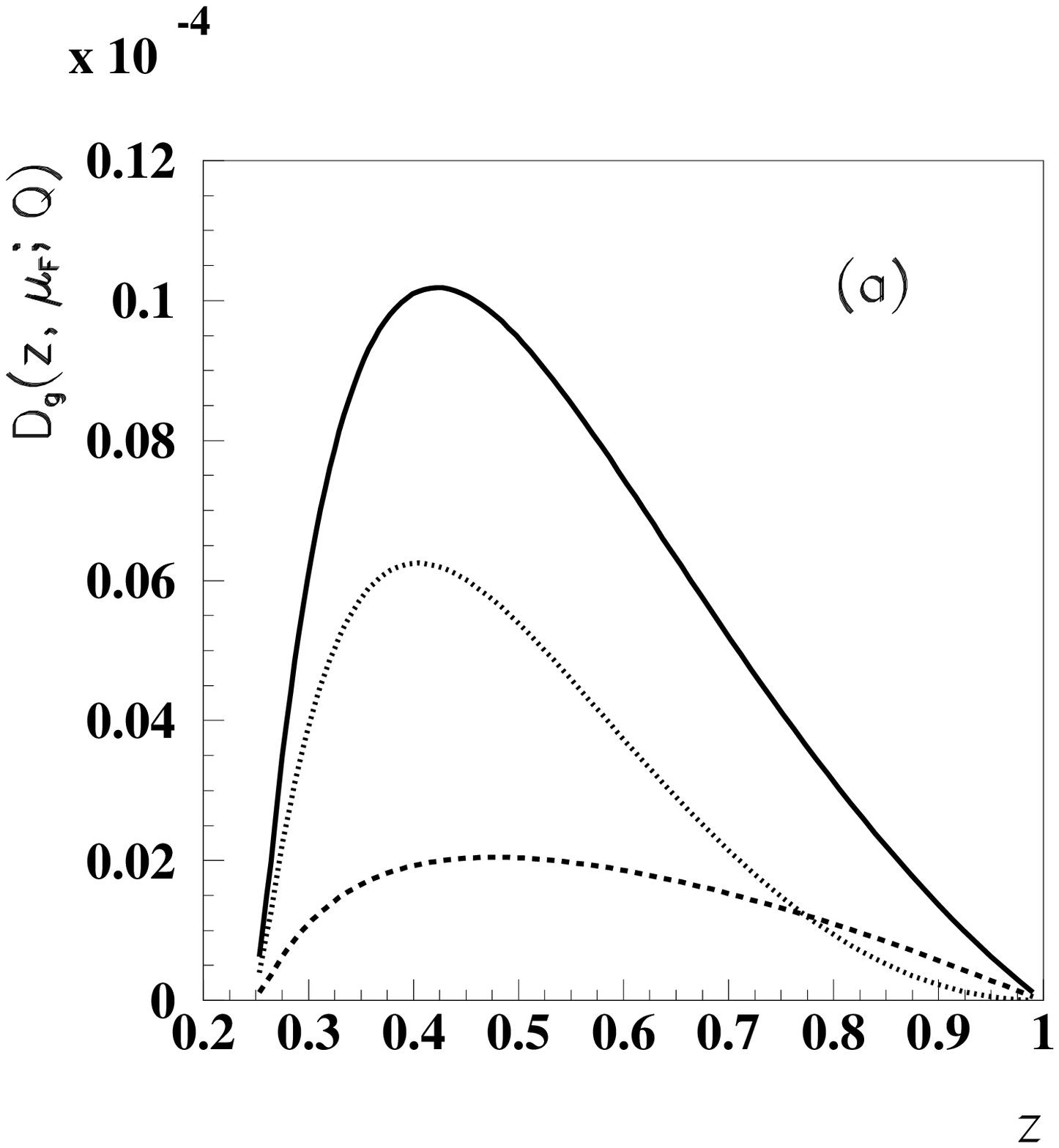,width=3.0in}
\end{minipage}
\hfil
\begin{minipage}[t]{3in}
\epsfig{figure=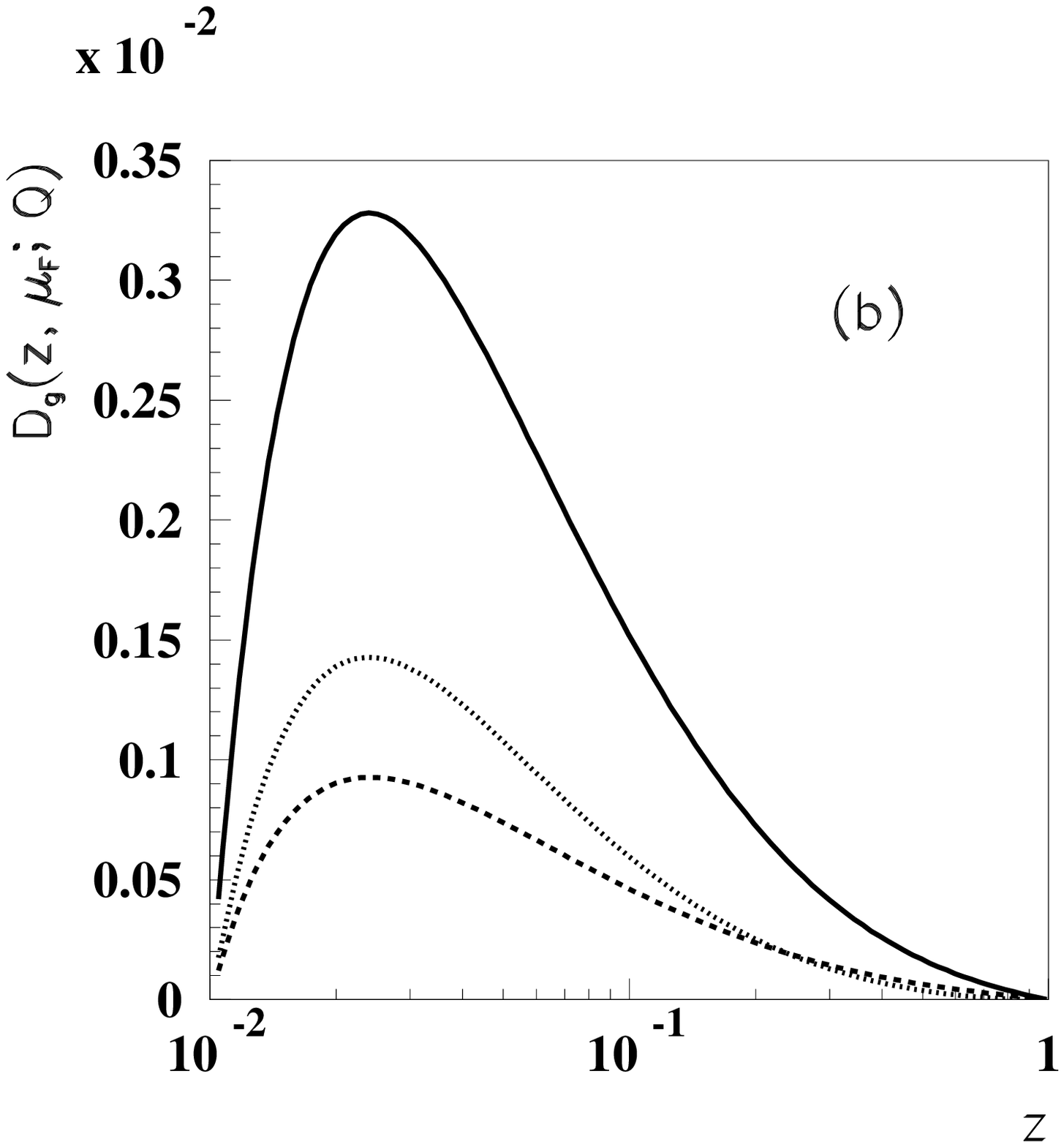,width=3.0in}
\end{minipage}
\caption{QCD generated gluon-to-virtual-photon fragmentation functions
as a function of $z$ at $Q=5$~GeV and $\mu_F=10$~GeV (a), and
$\mu_F=50$~GeV (b).  The solid, dashed, and dotted lines are for
unpolarized, transversely polarized, and longitudinally polarized
virtual photons, respectively. } 
\label{fig5}
\end{figure}

Although we do not have the lowest order gluon-to-virtual-photon
fragmentation functions due to
$\gamma_{g\rightarrow\gamma^*}^{(0)}=0$, QCD evolution in
Eq.~(\ref{RG-unpol}) can generate the gluon-to-virtual-photon
fragmentation functions.  In Figs.~\ref{fig5}(a) and \ref{fig5}(b), we 
plot gluon-to-virtual-photon fragmentation functions at $\mu_F=10$~GeV
and $\mu_F=50$~GeV, respectively.  The virtual photon mass is again
chosen to be $Q=5$~GeV.  As shown in Fig.~\ref{fig5}, QCD evolution
generated gluon-to-virtual-photon fragmentation functions at $Q=5$~GeV
grow very fast when the fragmentation scale $\mu_F^2$ increases. 
They are about two orders of magnitude smaller than the
quark-to-virtual-photon fragmentation functions at $\mu_F=10$~GeV, and
only one order of magnitude smaller at $\mu_F=50$~GeV.  Therefore, at 
$Q=1.5$~GeV or at a larger value of $\mu_F$, QCD generated
gluon-to-virtual-photon fragmentation functions become more
important. 

In conclusion, we have argued that virtual photon fragmentation
functions are well-defined and physically meaningful.  We derive the
evolution equations for virtual photon fragmentation functions, and
show that these fragmentation functions are perturbatively calculable.
We demonstrate that QCD resummation of the large logarithms 
caused by quark and gluon radiation provides a very important
contribution to the fragmentation functions when the fragmentation
scale $\mu_F$ is large and/or the invariant mass $Q$ is relatively
small.  Just like pion fragmentation functions \cite{Owens-RMP}, the
virtual photon fragmentation functions derived here are universal and
can be applied to any processes with massive lepton-pair production.
In the rest of this paper, we discuss some potential applications of
our results.
   
Contributions of fragmentation functions $D(z)$ to the physical
observables generally depend on the production of the parent partons 
at the same $z$.  Since the cross sections for producing the parent
partons strongly depend on the momenta of the produced partons, the 
role of the fragmentation functions may be different for different
regions of the $z$ values.  For example, for the Drell-Yan production,
if the cross section is dominated by the small $z$ region, the
produced virtual photons will likely be longitudinally polarized.  On
the other hand, if the cross section is dominated by the large $z$
region, the virtual photon will be transversely polarized.  For a
fixed collision energy $\sqrt{S}$, the Drell-Yan cross sections depend
on the $z$-value from $z_{min} = \sqrt{(Q^2+Q_T^2)/S}\, 
[ {\rm e}^{y} + {\rm e}^{-y} ]$ to 1 with the rapidity $y$.  For
$Q=5$~GeV and $y=0$ at $\sqrt{S}=2$~TeV (the new Tevatron energy),
$z_{min}\approx 0.01$ and 0.05 for $Q_T=10$ and 50~GeV, respectively.
It is clear from Fig.~\ref{fig3}(b) that the produced virtual photons
at $Q_T\sim 50$~GeV are more likely to be transversely polarized.  At
$Q_T=10$~GeV, the longitudinally polarized fragmentation functions in
Fig.~\ref{fig3}(a) are much larger than the transverse fragmentation
functions in small $z$ region, and the $1/z^2$ factor in
Eq.~(\ref{q-fac}) for the convolution over $z$ is also favor for
producing a longitudinally polarized virtual photon. But, the cross
section for producing a parent parton of momentum $k_T$ at small
$z\approx Q_T/k_T$ is a very steep falling function of $k_T$, and
consequently, it strongly reduces the rate for producing a
longitudinally polarized virtual photon.  In conclusion, when $Q_T$ is
large, the virtual photon produced in Drell-Yan massive lepton-pair
production is more likely to be transversely polarized.

\begin{figure}
\begin{center}
\epsfig{figure=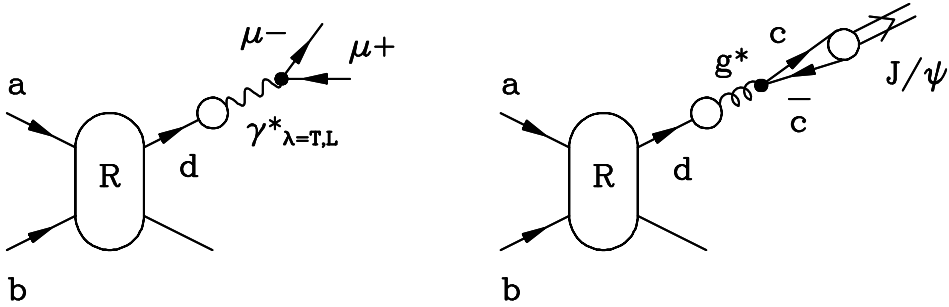,width=5.0in}
\end{center}
\caption{Sketch for Drell-Yan massive lepton-pair and J/$\psi$
production via parton fragmentation.}
\label{fig6}
\end{figure}

Recently, it was found that the J/$\psi$ mesons produced at Fermilab
Tevatron become more longitudinally polarized as the transverse
momentum increases \cite{CDF-jpsi}.  On the other hand, various
theoretical calculations predict the J/$\psi$ mesons to be more
transversely polarized \cite{BKL-jpsi}. The virtual photon production
(extracted from Drell-Yan massive lepton-pair production) at large
$Q_T$ and small $Q^2$ has a lot in common with the J/$\psi$ production
at high $Q_T$.  They both have two large physical 
scales: $Q_T$ and $Q^2$, which is equal to $M_{{\rm J/}\psi}^2$ in the
case of J/$\psi$ production; and $Q_T^2$ is much larger than $Q^2$.  
If the collision energy $\sqrt{S}$ is large enough and the logarithm
$\ln(Q_T^2/Q^2)$ is so important that the resummed  
fragmentation contributions dominate the production cross sections,
the virtual photon and J/$\psi$ production will share the {\it same}
partonic subprocesses, labeled by the $R$ in Fig.~\ref{fig6}. 
With their respected fragmentation functions, both the virtual photon
and J/$\psi$ production at high $Q_T$ are perturbatively calculable
\cite{QS-jpsi}.  As discussed above, the virtual photon fragmentation
functions are completely perturbative, while the parton to J/$\psi$
fragmentation functions involve final-state nonperturbative
interactions.  Therefore, only difference between the virtual photon
and J/$\psi$ production at high $Q_T$ and large $\sqrt{S}$ is the
final-state strong interactions during the formation of J/$\psi$
meson.  We propose to measure the virtual photon polarization in
Drell-Yan massive lepton-pair production at large $Q_T$ and low $Q^2$. 
The measurements of the virtual photon's polarization at high $Q_T$
provide not only a good test of QCD perturbation theory, but also a
reference process to test the models of J/$\psi$ formation.  

As shown in Fig.~\ref{fig6}, the fragmentation functions from a parton
$d$ to a physical J/$\psi$ can be approximated by the fragmentation
functions to a virtual gluon of invariant mass $Q$, which immediately
decay into a $c\bar{c}$-pair, convoluted with a transition from the
$c\bar{c}$-pair to a physical J/$\psi$ \cite{QS-jpsi}.  The
fragmentation functions for a parton to a virtual gluon should be
perturbatively calculable, and their dependence on the polarization
should be very similar to that of the virtual photon fragmentation
functions.  But, the produced charm and anticharm quark pair of
invariant mass $Q$ can in principle radiate gluons and have soft
interactions with other partons in the collisions.  However, due to
the heavy quark mass, such final-state interactions during the
formation or transition from the produced $c\bar{c}$-pair to a
physical J/$\psi$ meson are not expected to significantly change the
polarization \cite{BKL-jpsi}.  If the formation from the
$c\bar{c}$-pair of invariant mass $Q$ to a physical J/$\psi$ meson
does not change the polarization, one can expect the polarization of
the J/$\psi$ mesons produced at high $Q_T$ to be similar to the
polarization of the virtual photon in Drell-Yan massive lepton-pair
production at the same kinematics.  For example, the non-relativistic
QCD (NRQCD) model of J/$\psi$ production predicts the J/$\psi$ mesons
to be transversely polarized at large transverse momentum
\cite{BKL-jpsi}, which is not consistent with recent  
Fermilab data \cite{CDF-jpsi}.  Since the leptons do not interact
strongly once produced, the measurements of the virtual photon
polarization in Drell-Yan massive lepton-pair production at large
transverse momentum should help us to narrow the questions about
J/$\psi$ production.


\section*{Acknowledgment}

We thank G. Sterman for very helpful discussions. 
This work was supported in part by the U.S. Department of Energy under
Grant No. DE-FG02-87ER40731.



\vskip 0.4in

\end{document}